\documentclass[12pt]{article}
\usepackage[dvips]{color}
\usepackage{amsmath}
\usepackage{graphicx}
\usepackage{subfigure}
\usepackage{epsfig}
\usepackage{amsmath}
\usepackage{cite}
\usepackage{color}
\usepackage{subfigure}

\begin{document}
\begin{center}
\Large{\bf Scalar Weak Gravity Conjecture in Super Yang-Mills Inflationary Model}\\
\small \vspace{1cm} {\bf Jafar Sadeghi$^{\star}$\footnote {Email:~~~pouriya@ipm.ir}}, \quad
{\bf Mohammad Reza Alipour $^{\star}$\footnote {Email:~~~mr.alipour@stu.umz.ac.ir}}, \quad
{\bf Saeed Noori Gashti$^{\star}$\footnote {Email:~~~saeed.noorigashti@stu.umz.ac.ir}}, \quad
\\
\vspace{0.5cm}$^{\star}${Department of Physics, Faculty of Basic
Sciences,\\
University of Mazandaran
P. O. Box 47416-95447, Babolsar, Iran}\\
\small \vspace{1cm}
\end{center}
\begin{abstract}
In this article, we want to check four inflation models, such as  composite NJL inflation (NJLI),   Glueball inflation(GI),  super Yang-Mills inflation (SYMI), and Orientifold inflation (OI), with two conjectures of the swampland program: scalar weak gravity conjecture (SWGC) and strong scalar weak gravity conjecture (SSWGC) since all these models violate the dS swampland conjecture(DSC) but are compatible with further refining de Sitter swampland conjecture (FRDSSC) through manual adjustment of free parameters of the mentioned conjecture. We want to study the simultaneous compatibility of each model with these two new conjectures. Despite being consistent with (FRDSSC), we find that all models are not compatible with the other conjectures of the Swampland program in all regions, and these conjectures are only satisfied in a specific area. Also, due to the presence of constant parameter $(\phi_{0})$ in the higher orders derivatives, the (SYMI) and (OI) among all the models are more compatible with all conjectures of the swampland program. These models can provide a more significant amount of satisfaction with all of them. They can be suitable and accurate inflation models for a more profound examination of universe developments. We determined a particular region for these models is compatible with (FRDSSC), (SWGC), and (SSWGC) simultaneously.
\end{abstract}
\newpage
\tableofcontents

\section{Introduction}
Swampland program has recently been presented to evaluate and prove string theory to investigate effective low energy theories related to quantum gravity.
In recent years, many efforts have been made to develop the theory of everything, and perhaps string theory is one of the most well-known theories in this route.
Since, the string theory is explained very clearly, we can expect various consequences from it in cosmology.
As a result, a wide range of ideas and structures in the literature takes an in-depth look at the cosmological implications of string theory\cite{b,e,i,t}.
In string theory, many possible vacuums are created, which are one of the exciting aspects of the theory's cosmology; the set of these also forms the string theory landscape.
The important question in this regard is which of the effective low energy theories can be compatible with the string theory\cite{b,e,i,t}.
Therefore, the swampland program was introduced. This program includes many conjectures, including weak gravity conjecture (WGC), dS and AdS conjectures, SWGC, SSWGC, TCC, etc\cite{a,b,c,d,e,f,h,i,j,k,l,m,n,o}.
The collection of effective low-energy theories compatible with quantum gravity lives in the landscape.
A broader area surrounds the landscape, and the set of theories incompatible with quantum gravity is placed in this area, referred to as the swampland.
Many researchers have accepted string theory as a theory that determines quantum gravity.
Therefore, effective low-energy theories compatible with these swampland conjectures are aligned with quantum gravity, which can be of great help in finding a solution to advance this great question that has occupied researchers for years, namely, quantum gravity. So far, much work has been examined about the swampland program in literature; you can see Ref.s\cite{a,b,c,d,e,f,h,i,j,k,l,m,n,o,p,q,r,s,t,tt,ttt,u,v,w,x,y,z,aa,bb,cc,dd,ee,ff,gg,hh,ii,jj,kk,ll} for further study.
The swampland program has played an important role in finding phenomena consistent with quantum gravity. It has been used in various parts of physics, including black holes, inflation, and dark energy. Recently, many improvements have been made in the swampland program, which can be used to solve many cosmological problems. Among them, we can mention the de Sitter and refined de Sitter swampland conjecture that it is possible to find models compatible with quantum gravity by using the derivative of scalar field potentials \cite{1,2}. One investigate the four-dimensional theory of a real field $\varphi^i$ coupled to gravity, whose dynamics can be controlled using a scalar potential $V(\varphi^j)$ and whose action is as follows \cite{2,3},
\begin{equation}\label{eq1}
S=\int_{4D} d^4x \sqrt{-g}\left[-\frac{1}{2}M_p^2 R+\frac{1}{2}g^{\mu\nu}h_{ij} \partial_{\mu}\varphi^i \partial_{\nu}\varphi^j-V\right]
\end{equation}
where $g$ is the matrix metric in the four-dimensional space, $M_p$ is the Planck mass, $R$ is the Riemann curvature in the four-dimensional space, and $h_{ij}$ is the metric of the field space. Therefore, we examine different phenomenological models, such as inflation, described by action \eqref{eq1}.
Among the conjectures of the swampland program used to investigate cosmology is the de Sitter and refined de Sitter swampland conjecture, which state that the effective theories of quantum gravity placed in the landscape must satisfy at least one of the following constraints \cite{4,5},
\begin{equation}\label{2}
|\nabla V|\geq\frac{c_{1}}{M_{p}}V, \hspace{12pt} min(\nabla_{i}\nabla_{j}V)\leq -\frac{c_{2}}{M_{pl}^{2}}V
\end{equation}
The above equations for the $V>0$ can be rewritten in terms of the slow-roll parameters as follows,
\begin{equation}\label{3}
\sqrt{2\epsilon_{V}}\geq c_{1} ,\hspace{12pt}  or \hspace{12pt}\eta_{V}\leq -c_{2}
\end{equation}
where $c_1$ and $c_2$ are both positive and order of one, i.e., $c_1=c_2=\mathcal{O}(1)$. Also, the left side of equation (2) is related to the main swampland conjecture. Recently, David Andriot and Christoph Roupec combined the two of the de Sitter swampland conjecture and refined and formulated them, which is called further refining de Sitter swampland conjecture. It states that an effective low-energy theory of quantum gravity that consider the action equation(1) must satisfy the following relation \cite{2,3},
\begin{equation}\label{4}
\bigg(M_{p}\frac{|\nabla V|}{V}\bigg)^{q}-aM_{p}^{2}\frac{min(\nabla_{i}\nabla_{j}V)}{V}\geq b,
\end{equation}
where $a+b=1$, $a,b>0$, $q>2$. $a$, $b$, and $q$ are its free constant parameters that create a restriction for this conjecture. Many inflationary models have been investigated using this conjecture. The advantage of this conjecture over the old conjecture is that, unlike the refined dS conjecture, it doesn't sound inconsistent with the slow-roll single-field inflationary models \cite{3}.
One of the other important conjectures of the swampland program is the weak gravity conjecture $WGC$, which states that gravity is the weakest force.
Also, Palti generalized the WGC and showed the scalar field forces are stronger than gravity \cite{6,7}. Considering a particle $h$ with mass $m$ coupled to a light scalar $\varphi$ whose mass is a function of the scalar, in that case, the scalar weak gravity conjecture ($SWGC$) states that the intermediate force is stronger than gravity, and assuming $m^2=V^{\prime \prime}=\frac{\partial^2 V}{\partial \varphi^2}$, we have the following condition for SWGC,
\begin{equation}\label{eq5}
(V^{(3)})^2\geq \frac{(V^{(2)})^2}{M_p^2},
\end{equation}
where the power number in the parentheses means the order of the derivative relative to $\varphi$. Also, Eduardo Gonzalo and Luis E.Ibáñez suggested a strong version of SWGC, i.e., SSWGC \cite{8} which expresses that the potential of any canonically normalized real scalar $V(\varphi)$ must satisfy any value of the field restriction:
\begin{equation}\label{eq6}
2(V^{(3)})^2 - V^{(2)}V^{(4)}\geq \frac{(V^{(2)})^2}{M_p^2}.
\end{equation}
In this article, we are trying to test the considered inflation models with $SWGC$ and $SSWGC$ conjectures to find the model compatible with quantum gravity.
Therefore, according to all the above explanations, we organize the article.\\
In section 2, we overview the inflation models such as (NJLI), (GI), (SYMI) and (OI) in 4 subsections.
In section 3, we challenge these inflationary models with two conjectures of the swampland program, i.e., the SWGC and SSWGC. We will discuss the compatibility or incompatibility of each model with mentioned conjecture and determine the consistent regions. We compare the results with each other. Finally, we describe the outcomes in Section 4.

\section{Overview of inflationary models}
In this section, according to \cite{a}, we briefly introduce four inflation models composite NJL Inflation(NJLI),   Glueball inflation(GI),  super Yang-Mills inflation (SYMI), and Orientifold inflation (OI). We review the results of each model's compatibility with the two swampland conjectures described in \cite{a}. Then, in the next section, using the potential of each model, we will check other important conjectures of the swampland program, namely (SWG) and (SSWG). Finally, we will introduce the results thoroughly, and the best model that has the most compatibility with all conjectures presented as well as the best inflation model to examine the universe development.
\subsection{Model I: NJLI}
The action expressing the inflationary model that the inflation has a non-minimally coupling with gravity is defined as follows in Jordan's framework \cite{a,100},
\begin{equation}\label{7}
\begin{split}
&S_{J}=\int d^{4}x\sqrt{-g}\bigg(-\frac{1}{2}M_{p}^{2}R+\frac{1}{2}g^{\mu\nu}\partial_{\mu}\varphi\partial_{\nu}\varphi-\frac{\xi R}{2}\Big[\varphi^{2}-\frac{\upsilon^{2}}{2}\Big]-V_{J}(\varphi)\bigg)\\
&V_{J}(\varphi)=-\frac{1}{2}m^{2}_{\varphi}\varphi^{2}+\frac{1}{2}\lambda\varphi^{4},
\end{split}
\end{equation}
where ($\upsilon$) and ($\varphi$) specify vacuum expectation value and inflation field. Also, the index (J) shows the Jordan frame. The action, as mentioned earlier, can be transformed into an Einstein framework by applying conformal transformation with a new canonical normalized field as a minimally coupled form. Hence, this conformal transformation is expressed in the following form \cite{100,101},
\begin{equation}\label{8}
\begin{split}
\widetilde{g}_{\mu\nu}=\Omega^{2}g_{\mu\nu}=\bigg(1+\frac{\xi(\varphi^{2}-\upsilon^{2}/2)}{M_{p}^{2}}\bigg)g_{\mu\nu}.
\end{split}
\end{equation}
The action in equation (7) is rewritten in Einstein's framework, where the index (E) is the characteristic of Einstein's frame,
\begin{equation}\label{9}
\begin{split}
S_{E}=\int d^{4}x\sqrt{-g}\bigg(-\frac{1}{2}M_{p}^{2}R+\frac{1}{2}\Omega^{-4}\Big(\Omega^{2}+\frac{6\xi\varphi^{2}}{M_{p}^{2}}\Big)g^{\mu\nu}\partial_{\mu}\phi\partial_{\nu}\phi-U(\varphi)\bigg)\\
\end{split}
\end{equation}
where
\begin{equation}\label{10}
\begin{split}
&\Omega^{2}=\bigg(1+\frac{\xi(\varphi^{2}-\upsilon^{2}/2)}{M_{p}^{2}}\bigg)\\
&U(\varphi)\equiv\Omega^{-4}V_{J}(\varphi)
\end{split}
\end{equation}
According to \cite{a} by introducing a new canonically normalized scalar field, we will have,
\begin{equation}\label{11}
\begin{split}
\frac{1}{2}g^{\mu\nu}\partial_{\mu}\chi(\varphi)\partial_{\nu}\chi(\varphi)=\frac{1}{2}\big(\frac{d\chi}{d\varphi}\big)^{2}g^{\mu\nu}\partial_{\mu}\varphi\partial_{\nu}\varphi,
\end{split}
\end{equation}
where
\begin{equation}\label{12}
\begin{split}
\big(\frac{d\chi}{d\varphi}\big)=\sqrt{\Omega^{-4}\Big(\Omega^{2}+\frac{6\xi\varphi^{2}}{M_{p}^{2}}\Big)}.
\end{split}
\end{equation}
In the limit $\xi\varphi^{2}\ll M_{p}^{2}$, that is, the small field values, the potential for the new field becomes the original field, which is not valid for the $\xi\varphi^{2}\gg M_{p}^{2}$. Therefore, the field solution ($\varphi$) is rewritten according to the new field $\chi$ in the following form \cite{a,100},
\begin{equation}\label{13}
\begin{split}
\varphi\simeq\frac{M_{p}}{\sqrt{\xi}}\exp\Big(\frac{\chi}{\sqrt{6}M_{p}}\Big).
\end{split}
\end{equation}
Thus, the effective potential is also expressed as follows,
\begin{equation}\label{14}
\begin{split}
U(\chi)\simeq\frac{\lambda M_{p}^{4}}{2\xi^{2}}\bigg(1+\exp\Big[-\frac{2\chi}{\sqrt{6}M_{p}}\Big]\bigg)^{-2}.
\end{split}
\end{equation}
The authors in \cite{a} challenged this inflationary model according to one of the conjectures of the swampland program. It was found that the model is in strong tension with dS swampland conjectures because $C_{1}=C_{2}\neq\mathcal{O}(1)$\cite{a}. Therefore, they checked the model with another conjecture: further refining dS swampland  conjecture. By manually adjusting the parameters of the mentioned conjecture, namely a, b and q, they showed that the mentioned model is compatible with it. In the next section, we will examine this model with other conjectures of the swampland program,i.e., (SWGC) and (SSWGC), and explain the results in detail.

\subsection{Model II: GI}
We provide a brief description of this model. According to \cite{a}, the action of the corresponding model is expressed in the following form, where the model has a general non-minimal coupling to gravity \cite{102,103},
\begin{equation}\label{15}
\begin{split}
S=\int d^{4}x\sqrt{-g}\bigg(-\frac{M_{p}^{2}+\xi\Lambda^{2}(\phi/\phi_{0})^{2}}{2}R+L_{GI}\bigg),
\end{split}
\end{equation}
where,
\begin{equation}\label{16}
\begin{split}
&L_{GI}=\varphi^{-3/2}\partial_{\mu}\varphi\partial^{\mu}\varphi-\frac{\varphi}{2}\ln\big(\frac{\varphi}{\Lambda^{4}}\big)\\
&\frac{\varphi}{\Lambda^{4}}=\big(\frac{\phi}{\phi_{0}}\big)^{4}\\
&\phi_{0}=4\sqrt{2}\Lambda,
\end{split}
\end{equation}
which $\Lambda$ is called mass scale, and parameter $\xi$ characterized the coupling to gravity, respectively. According to the above explanations, the action in Einstein's frame is rewritten as follows \cite{a,102,103},
\begin{equation}\label{17}
\begin{split}
S=\int d^{4}x\sqrt{-g}\bigg(-\frac{1}{2}M_{p}^{2}R+\Omega^{-2}\bigg[1+\frac{3\xi^{2}\Lambda^{2}(\phi/\phi_{0})^{2}}{16M_{p}^{2}}\Omega^{-2}\bigg]\big(\frac{\Lambda}{\phi_{0}}\big)^{2}\partial_{\mu}\phi\partial^{\mu}\phi-\Omega^{-4}V_{GI}\bigg)
\end{split}
\end{equation}
where $\Omega^{2}=\big(M_{P}^{2}+\xi\Lambda^{2}(\phi/\phi_{0})^{2}\big)/M_{p}^{2}$. According to $\xi\neq 0$ and the large field limit, the mentioned equation is reduced to $\Omega^{2}\simeq \xi\Lambda^{2}(\phi/\phi_{0})^{2}/M_{p}^{2}$, and the potential is calculated \cite{a,102,103},
\begin{equation}\label{18}
\begin{split}
V_{GI}=2\Lambda^{4}(\phi/\phi_{0})^{4}\ln(\phi/\phi_{0})
\end{split}
\end{equation}
where $\phi_{0}\equiv4\sqrt{2}\Lambda$. According to the definition, we will consider a canonically normalized field $\chi$ associated with ($\phi$),
\begin{equation}\label{19}
\begin{split}
\frac{1}{2}\widetilde{g}^{\mu\nu}\partial_{\mu}\chi(\phi)\partial_{\nu}\chi(\phi)=\frac{1}{2}\big(\frac{d\chi}{d\phi}\big)^{2}\widetilde{g}^{\mu\nu}\partial_{\mu}\phi\partial_{\nu}\phi
\end{split}
\end{equation}
where
\begin{equation}\label{20}
\begin{split}
\frac{1}{2}\big(\frac{d\chi}{d\phi}\big)^{2}=\Omega^{-2}\bigg[1+\frac{3\xi^{2}\Lambda^{2}(\phi/\phi_{0})^{2}}{16M_{p}^{2}}\Omega^{-2}\bigg]\big(\frac{\Lambda}{\phi_{0}}\big)^{2}
\end{split}
\end{equation}
By considering a condition as $\chi\propto \ln \phi$, the potential reduces to $\Omega^{-4}V_{GI}\propto \ln \phi$ \cite{a,102,103}. Hence, we will have the following equation with respect to the canonically normalized field,
\begin{equation}\label{21}
\begin{split}
S_{E}=\int d^{4}x\sqrt{-g}\bigg(-\frac{1}{2}M_{p}^{2}R+\frac{1}{2}g^{\mu\nu}\partial_{\mu}\chi\partial_{\nu}\chi-U_{GI}(\chi)\bigg),
\end{split}
\end{equation}
where
\begin{equation}\label{22}
\begin{split}
U_{GI}(\chi)=\Omega^{-4}V_{GI}(\varphi).
\end{split}
\end{equation}
The potential of this model in Einstein's frame is rewritten in terms of field ($\phi$) according to \cite{a},  slow-roll analysis and the large field regime $N_{c}^{2}\xi\Lambda^{2}(\phi/\phi_{0})^{2}\gg M_{p}^{2}$,
\begin{equation}\label{23}
\begin{split}
U_{GI}(\phi)=\frac{2M_{p}^{4}}{\xi^{2}}\ln\big(\frac{\phi}{\phi_{0}}\big)
\end{split}
\end{equation}
Similar to the previous situation, the authors in\cite{a} concluded that this model is incompatible with the dS conjecture and does not satisfy it. Then they applied FRDSSC to this model and proved complete compatibility between the model and this conjecture by manually adjusting the free parameters. We will also challenge this model with two other conjectures of the swampland program and describe the results in detail.

\subsection{Model III: SYMI}
We also briefly explain this model, which you can see \cite{a} for further study. The action in Jordan's frame for this model is generally expressed in the following form according to the non-minimally coupled gravity structure \cite{102},
\begin{equation}\label{24}
\begin{split}
S_{J}=\int d^{4}x\sqrt{-g}\bigg(-\frac{M^{2}+N_{c}^{2}\xi\Lambda^{2}(\phi/\phi_{0})^{2}}{2}R+L_{SYM}\bigg)
\end{split}
\end{equation}
where
\begin{equation}\label{25}
\begin{split}
&L_{SYM}=-\frac{N_{c}^{2}}{\alpha}(\varphi\varphi^{\dagger})^{-2/3}\partial_{\mu}\varphi\partial^{\mu}\varphi^{\dagger}-\frac{4\alpha N_{c}^{2}}{9}(\varphi\varphi^{\dagger})^{2/3}\ln\big(\frac{\varphi}{\Lambda^{3}}\big)\ln\big(\frac{\varphi^{\dagger}}{\Lambda^{3}}\big)\\
&\frac{\varphi}{\Lambda^{3}}=\big(\frac{\phi}{\phi_{0}}\big)^{3}\\
&\phi_{0}=3N_{c}\big(\frac{2}{\alpha}\big)^{1/2}\Lambda .
\end{split}
\end{equation}
In the above equations, $\alpha$ is a constant parameter, and $\Lambda$ is the mass scale. We also focus on the real part of the inflation field ie ($\varphi=\varphi^{\dagger}$) \cite{a,102}. The above action is rewritten in Einstein's framework \cite{102},
\begin{equation}\label{26}
\begin{split}
&S_{E}=\int d^{4}x\sqrt{-g}\bigg(-\frac{1}{2}M_{p}^{2}R+\frac{9N_{c}^{2}}{\alpha}\Omega^{-2}\bigg[1+\frac{\alpha N_{c}^{2}\xi^{2}}{3M_{p}^{2}}\Omega^{-2}\Lambda^{2}\big(\frac{\phi}{\phi_{0}}\big)^{2}\bigg]\\
&\times\big(\frac{\Lambda}{\phi_{0}}\big)^{2}\partial_{\mu}\phi\partial^{\mu}\phi-\Omega^{-4}V_{SYM}\bigg)
\end{split}
\end{equation}
Assuming $\xi\neq 0$, $\Omega^{2}\simeq N_{c}^{2}\xi\Lambda^{2}(\phi/\phi_{0})^{2}/M_{p}^{2}$ and according to the explanations in\cite{a}, we will have
\begin{equation}\label{27}
\begin{split}
V_{SYM}(\phi)=4\alpha N_{c}^{2}\Lambda^{4}\big(\frac{\phi}{\phi_{0}}\big)^{4}\ln^{2}\big(\frac{\phi}{\phi_{0}}\big)
\end{split}
\end{equation}
By introducing a canonically normalized field related to the $\phi$ through the following relations \cite{a},
\begin{equation}\label{28}
\begin{split}
\frac{1}{2}\widetilde{g}^{\mu\nu}\partial_{\mu}\chi(\phi)\partial_{\nu}\chi(\phi)=\frac{1}{2}\big(\frac{d\chi}{d\phi}\big)^{2}\widetilde{g}^{\mu\nu}\partial_{\mu}\phi\partial_{\nu}\phi
\end{split}
\end{equation}
where
\begin{equation}\label{29}
\begin{split}
\frac{1}{2}\big(\frac{d\chi}{d\phi}\big)^{2}=\frac{9N_{c}^{2}}{\alpha}\Omega^{-2}\bigg[1+\frac{\alpha N_{c}^{2}\xi^{2}}{3M_{p}^{2}}\Omega^{-2}\Lambda^{2}\big(\frac{\phi}{\phi_{0}}\big)^{2}\bigg]\big(\frac{\Lambda}{\phi_{0}}\big)^{2}
\end{split}
\end{equation}
Thus, we can rewrite the action in Einstein's frame in terms of the canonically normalized field as follows,
\begin{equation}\label{30}
\begin{split}
S_{E}=\int d^{4}x\sqrt{-g}\bigg(-\frac{1}{2}M_{p}^{2}R+\frac{1}{2}g^{\mu\nu}\partial_{\mu}\chi\partial_{\nu}\chi-U_{SYM}(\chi)\bigg)
\end{split}
\end{equation}
where
\begin{equation}\label{31}
\begin{split}
U_{SYM}(\chi)=\Omega^{-4}V_{SYM}(\chi)
\end{split}
\end{equation}
According to \cite{a} and slow-roll analysis of the potential, also taking into account the large field regime $N_{c}^{2}\xi \Lambda^{2}\big(\phi/\phi_{0}\big)^{2}\gg M_{p}$, the model's potential in Einstein's frame is calculated in terms of the field $\phi$ \cite{104},
\begin{equation}\label{32}
\begin{split}
U_{SYM}(\phi)=\frac{4\alpha}{N_{c}^{2}}\frac{M_{p}^{4}}{\xi^{2}}\ln^{2}\big(\frac{\phi}{\phi_{0}}\big).
\end{split}
\end{equation}
The authors showed in \cite{a} that this model, similar to the previous models, violates the dS swampland conjecture and agrees and is compatible with FRDSSC. Therefore, in this article, we challenge this model concerning other conjectures of the swampland. So we can introduce the best model that is more compatible with all conjectures of the swampland program.

\subsection{Model IV: OI}
To introduce the final model, we start by expressing the action in the Jordan frame by considering non-minimal coupled to gravity; hence we have \cite{102},
\begin{equation}\label{33}
\begin{split}
S_{J}=\int d^{4}x\sqrt{-g}\bigg(-\frac{M^{2}+N_{c}^{2}\xi\Lambda^{2}\big(\frac{\phi}{\phi_{0}}\big)^{2}}{2}R+L_{OI}\bigg)
\end{split}
\end{equation}
where
\begin{equation}\label{34}
\begin{split}
&L_{OI}=-\frac{N_{c}^{2}}{\alpha_{OI}}(\varphi\varphi^{\dagger})^{-2/3}\partial_{\mu}\varphi\partial^{\mu}\varphi^{\dagger}-\frac{4\alpha_{OI}N_{c}^{2}}{9}(\varphi\varphi^{\dagger})^{2/3}\bigg[\ln\big(\frac{\varphi}{\Lambda^{3}}\big)\ln\big(\frac{\varphi^{\dagger}}{\Lambda^{3}}\big)-\beta\bigg]\\
&\frac{\varphi}{\Lambda^{3}}=\big(\frac{\phi}{\phi_{0}}\big)^{3}\\
&\phi_{0}=3N_{c}\big(\frac{2}{\alpha}\big)^{1/2}\Lambda
\end{split}
\end{equation}
where M is a mass scale, $\beta=\mathcal{O}(1/N_{c})$ and $\varphi=\varphi^{\dagger}$, the above action in Einstein's frame is in the following form.
\begin{equation}\label{35}
\begin{split}
&S_{E}=\int d^{4}x\sqrt{-g}\bigg[-\frac{1}{2}M_{p}^{2}R+\frac{9N_{c}^{2}}{\alpha}\Omega^{-2}\bigg(1+\frac{\alpha N_{c}^{2}\xi^{2}}{3M_{p}^{2}}\Omega^{-2}\Lambda^{2}\big(\frac{\phi}{\phi_{0}}\big)^{2}\bigg)\big(\frac{\Lambda}{\phi_{0}}\big)^{2}\\
&\times\partial_{\mu}\phi\partial^{\mu}\phi-\Omega^{-4}V_{OI}\bigg]
\end{split}
\end{equation}
Here, taking into account conditions such as $\xi\neq 0$ and with respect to\cite{a}, we will have,
\begin{equation}\label{36}
\begin{split}
V_{OI}(\phi)=4\alpha N_{c}^{2}\Lambda^{4}\big(\frac{\phi}{\phi_{0}}\big)^{4}\bigg[\ln^{2}(\frac{\phi}{\phi_{0}}\big)-\frac{\beta}{9}\bigg]
\end{split}
\end{equation}
It is possible to introduce a canonically normalized field related to the $\phi$ with respect to the following equation,
\begin{equation}\label{37}
\begin{split}
\frac{1}{2}\widetilde{g}^{\mu\nu}\partial_{\mu}\chi(\phi)\partial_{\nu}\chi(\phi)=\frac{1}{2}\big(\frac{d\chi}{d\phi}\big)^{2}\widetilde{g}^{\mu\nu}\partial_{\mu}\phi\partial_{\nu}\phi,
\end{split}
\end{equation}
where
\begin{equation}\label{38}
\begin{split}
\frac{1}{2}\big(\frac{d\chi}{d\phi}\big)^{2}=\frac{9N_{c}^{2}}{\alpha}\Omega^{-2}\bigg(1+\frac{\alpha N_{c}^{2}\xi^{2}}{3M_{p}^{2}}\Omega^{-2}\Lambda^{2}\big(\frac{\phi}{\phi_{0}}\big)^{2}\bigg)\big(\frac{\Lambda}{\phi_{0}}\big)^{2}.
\end{split}
\end{equation}
According to the canonically normalized field, we have,
\begin{equation}\label{39}
\begin{split}
S_{E}=\int d^{4}x\sqrt{-g}\bigg(-\frac{1}{2}M_{p}^{2}R+\frac{1}{2}g^{\mu\nu}\partial_{\mu}\chi\partial_{\nu}\chi-U_{OI}(\chi)\bigg)
\end{split}
\end{equation}
where
\begin{equation}\label{40}
\begin{split}
U_{OI}(\chi)=\Omega^{-4}V_{OI}(\varphi)
\end{split}
\end{equation}
Like the previous models, assuming conditions such as slow-roll analysis of the potential, the large field regime $N_{c}^{2}\xi\Lambda^{2}\big(\phi/\phi_{0}\big)^{2}\gg M^{2}$, and according to \cite{a}, the final model potential is also calculated in the Einstein frame as follows,
\begin{equation}\label{41}
\begin{split}
U_{OI}(\phi)=\frac{4\alpha}{N_{c}^{2}}\frac{M_{P}^{4}}{\xi^{2}}\bigg[\ln^{2}\big(\frac{\phi}{\phi_{0}}\big)-\frac{\beta}{9}\bigg]
\end{split}
\end{equation}
Like the previous models, this final model satisfies the FRDSSC while it is inconsistent with the dS swampland conjecture. Hence, we face four models in different structures, all of which violate the dS swampland conjecture and satisfy FRDSSC. Therefore, we challenge all these models with other swampland conjectures to introduce a model compatible with all conjectures of the swampland, which can be considered a suitable model for further investigations of universe evolutions.

\section{SWGC and SSWGC on inflation Models}
In this section, we apply the SWGC and SSWGC of the swampland program to the mentioned inflation models. Considering that all mentioned models satisfy the FRDSSC, we intend to choose the best model with the highest compatibility with all of these swampland conjectures through further investigation. We will also explain the results in detail.

\subsection{Model I}
According to the SWGC and SSWGC in equations (5), (6) and using the potential of Model I (NJLI) in equation (14), we will have, we put $M_p=1$,
\begin{equation}\label{eq42}
U^{(1)}(\chi)\simeq\frac{2\lambda}{\sqrt{6}\xi^2}\frac{e^{\frac{2\chi}{\sqrt{6}}}}{(1+e^{\frac{2\chi}{\sqrt{6}}})^3},
\end{equation}
\begin{equation}\label{eq43}
U^{(2)}(\chi)\simeq \frac{-2 \lambda}{3\xi^2}\frac{e^{\frac{4}{\sqrt{6}}}\chi}{(1+e^{\frac{2\chi}{\sqrt{6}}})^4}{(-2+e^{\frac{2\chi}{\sqrt{6}}})},
\end{equation}
\begin{equation}\label{eq44}
U^{(3)}(\chi)\simeq\frac{4\lambda}{3\sqrt{6}\xi^2}\frac{e^{\frac{4}{\sqrt{6}}\chi}(4-7e^{\frac{2\chi}{\sqrt{6}}}+
e^{\frac{4\chi}{\sqrt{6}}})}{(1+e^{\frac{2\chi}{\sqrt{6}}})^5},
\end{equation}
\begin{equation}\label{eq45}
U^{(4)}(\chi)\simeq \frac{-8 \lambda}{9\xi^2}\frac{e^{\frac{4}{\sqrt{6}}\chi}(-8+33e^{\frac{2\chi}{\sqrt{6}}}-18e^{\frac{4\chi}{\sqrt{6}}}
+e^{\frac{6\chi}{\sqrt{6}}})}{(1+e^{\frac{2\chi}{\sqrt{6}}})^6}.
\end{equation}
Now, we put the above equation in the $SWGC$ equation(5). So, we get the following relationship,
\begin{equation}\label{eq46}
\frac{4 \lambda^2 e^{\frac{8\chi}{\sqrt{6}}}}{9\xi^4(1+e^{\frac{2\chi}{\sqrt{6}}})^{10}}\left[
-(2+e^{\frac{2\chi}{\sqrt{6}}}-e^{\frac{4\chi}{\sqrt{6}}})^2+\frac{2}{3}(4-7e^{\frac{2\chi}{\sqrt{6}}}+e^{\frac{4\chi}{\sqrt{6}}})^2\right] \geq 0
\end{equation}
The above relation is greater than zero if the following condition is met,
\begin{equation}\label{eq47}
(\frac{2}{\sqrt{6}}+1)e^{\frac{4\chi}{\sqrt{6}}}-(\frac{14}{\sqrt{6}}+1)e^{\frac{2\chi}{\sqrt{6}}}+(\frac{8}{\sqrt{6}}+2) \geq 0
\end{equation}
Now, we use the change of variable $y=e^{\dfrac{2\chi}{\sqrt{6}}}$. A 2nd-degree equation is obtained in the following form,
\begin{equation}\label{eq48}
f(y)=(\frac{2}{\sqrt{6}}+1)y^2-(\frac{14}{\sqrt{6}}+1)y+(\frac{8}{\sqrt{6}}+2) \geq 0.
\end{equation}
First, we get the points where the function $f(y)=0$. We also calculate its minimum point. So, we will have,
\begin{equation}\label{eq49}
f(y)=0 \longrightarrow (y=1.129, y=2.568) \qquad \frac{\partial f(y)}{\partial y}=0 \longrightarrow y_{min}=1.849, f(y_{min})=-0.941 .
\end{equation}
Since, the minimum point is negative, $f(y)$ has a negative value in the interval $1.129<y<2.568$, so the SWGC is not satisfied in this range. We see the SWGC is met in the $\chi<0.148$ and $\chi>1.154$. Next, we check the SSWGC in equation (6). So, we will have,
\begin{equation}\label{eq50}
\frac{4\lambda^{2}\exp\big(\frac{8\chi}{\sqrt{6}}\big)}{27\xi^{4}(1+\exp\big(\frac{2\chi}{\sqrt{6}}\big))^{10}}\bigg[20-88\exp\big(\frac{2\chi}{\sqrt{6}}\big)+99\exp\big(\frac{4\chi}{\sqrt{6}}\big)-\exp\big(\frac{8\chi}{\sqrt{6}}\big)-10\exp\big(\sqrt{6}\chi\big)\bigg]\geq0.
\end{equation}
The above relationship is established if that, $$F(\chi)=\bigg[20-88\exp\big(\frac{2\chi}{\sqrt{6}}\big)+99\exp\big(\frac{4\chi}{\sqrt{6}}\big)-\exp\big(\frac{8\chi}{\sqrt{6}}\big)-10\exp\big(\sqrt{6}\chi\big)\bigg]\geq0.$$ As a result, we will plot a figure to check it.
\begin{figure}[hbt!]
\begin{center}
\includegraphics[width=.5\textwidth]{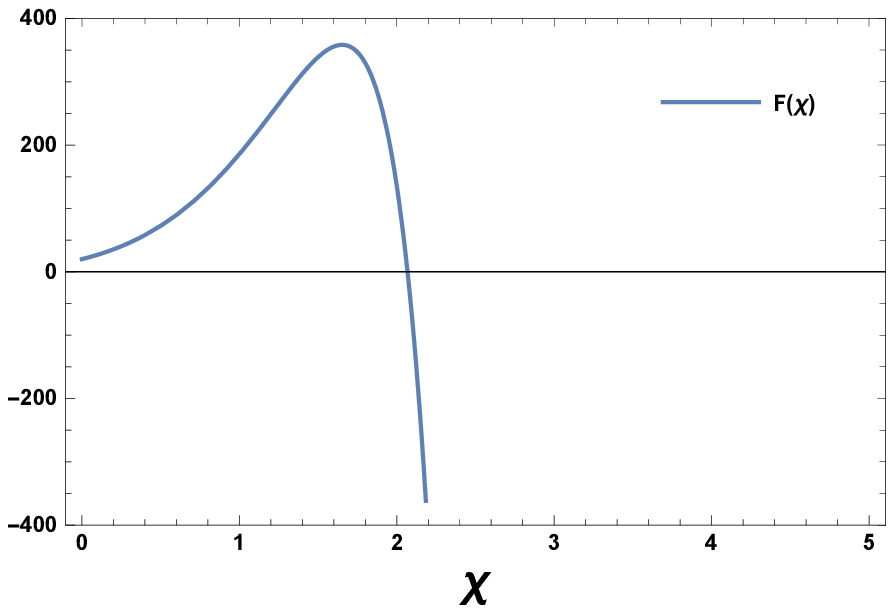}
\caption{$F(\chi)$} \label{fig:8a}
\end{center}
\end{figure}
From the Figure 1, we find that when $\chi \leq 2.068$, $F(\chi)\geq 0$, SSWGC is established.
Therefore, comparing the two conjectures, we can see that these two conjectures can be satisfied when $1.154\leq \chi \leq2.068$ and $\chi \leq0.148$.

\subsection{Model II}
We go through a similar process for all models to determine their compatibility with the mentioned conjectures. Therefore, with respect to the equations (5) and (6), also the potential of model II (GI) in equation(23), one can calculate,
\begin{equation}\label{eq51}
U_{GI}^{1}(\phi)=\frac{2}{\xi^{2}\phi}
\end{equation}
\begin{equation}\label{eq52}
U_{GI}^{2}(\phi)=-\frac{2}{\xi^{2}\phi^{2}}
\end{equation}
\begin{equation}\label{eq53}
U_{GI}^{3}(\phi)=\frac{4}{\xi^{2}\phi^{3}}
\end{equation}
\begin{equation}\label{eq54}
U_{GI}^{4}(\phi)=-\frac{12}{\xi^{2}\phi^{4}}
\end{equation}
Now, we first consider the SWGC. By putting the above relation in the equation(5), one can obtain
\begin{equation}\label{eq55}
\frac{4}{\xi^{2}\phi^{6}}(4-\phi^{2})\geq0
\end{equation}
According to the above relation, the SWGC is satisfied when $\phi<2$.
Next, we examine the SSWGC,
\begin{equation}\label{eq56}
\frac{4}{\xi^{2}\phi^{6}}(2-\phi^{2})\geq0
\end{equation}
According to the above relation, when $\phi<\sqrt{2}$, the SSWGC will be satisfied.
Therefore, both them (SWGC) and (SSWGC) are satisfied by sharing two conjectures, when $\phi<\sqrt{2}$.

\subsection{Model III}
We apply the conjectures to model III (SYMI), so with respect to equation (5), (6) and (32), we have
\begin{equation}\label{eq57}
U^1_{SYM}(\phi)=\frac{8\alpha}{N_c^2\xi^2\phi}(\ln(\frac{\phi}{\phi_0}))
\end{equation}
\begin{equation}\label{eq58}
U^{2}_{SYM}(\phi)=\frac{8\alpha}{N_{c}^{2}\xi^{2}\phi^{2}}\big(1-\ln\big(\frac{\phi}{\phi_{0}}\big)\big)
\end{equation}
\begin{equation}\label{eq59}
U^{3}_{SYM}(\phi)=\frac{8\alpha}{N_{c}^{2}\xi^{2}\phi^{3}}\big(-3+2\ln\big(\frac{\phi}{\phi_{0}}\big)\big)
\end{equation}
\begin{equation}\label{eq60}
U^{4}_{SYM}(\phi)=\frac{8\alpha}{N_{c}^{2}\xi^{2}\phi^{4}}\big(11-6\ln\big(\frac{\phi}{\phi_{0}}\big)\big)
\end{equation}
First, we consider SWGC. So, one can obtain,
\begin{equation}\label{eq61}
\frac{64\alpha^{2}}{N_{c}^{4}\xi^{4}\phi^{6}}F(\phi)\geq0,\hspace{12pt}F(\phi)=\bigg[\big(3-2\ln\big(\frac{\phi}{\phi_{0}}\big)\big)^{2}-\phi^{2}\big(1-\ln\big(\frac{\phi}{\phi_{0}}\big)\big)^{2}\bigg].
\end{equation}
Next, we discuss the SSWGC and reach the following equation,
\begin{equation}\label{eq62}
\frac{64\alpha^{2}}{N_{c}^{4}\xi^{4}\phi^{6}}G(\phi)\geq0,\hspace{12pt}G(\phi)=\bigg[7-\phi^{2}+(2\phi^{2}-7)\ln\big(\frac{\phi}{\phi_{0}}\big)+(2-\phi^{2})\big(\ln\big(\frac{\phi}{\phi_{0}}\big)\big)^{2}\bigg].
\end{equation}
In order for the two conjectures met, they must be $F(\phi)\geq 0$ and $G(\phi)\geq 0$. We need to find their common points to see in which interval of $\phi$ these conjectures are valid. Since, $F(\phi)$ and $G(\phi)$ are complex relations, we are trying to examine using the plot to find the points compatible with these conjectures.
\begin{figure}[hbt!]
\includegraphics[width=.28\textwidth]{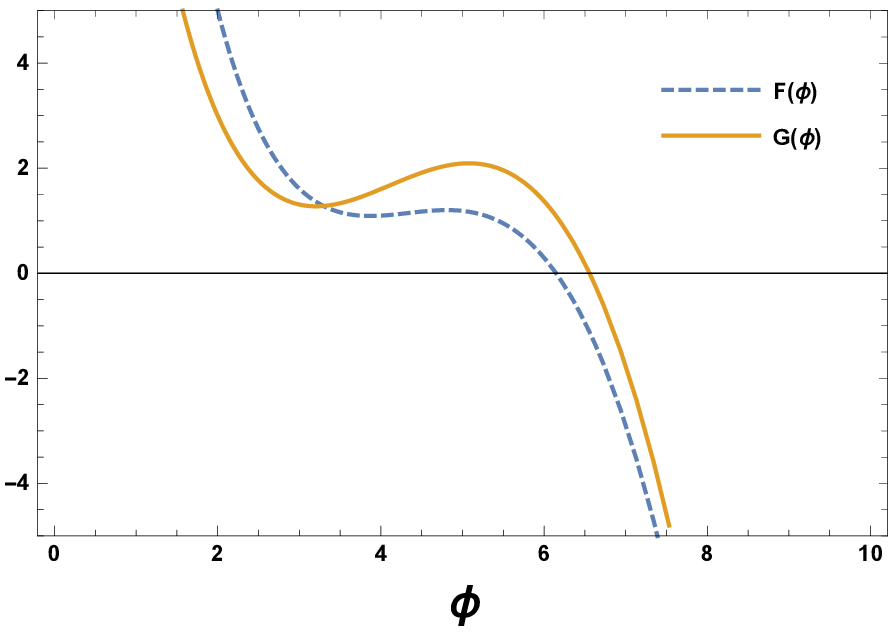}
(a)
\includegraphics[width=.28\textwidth]{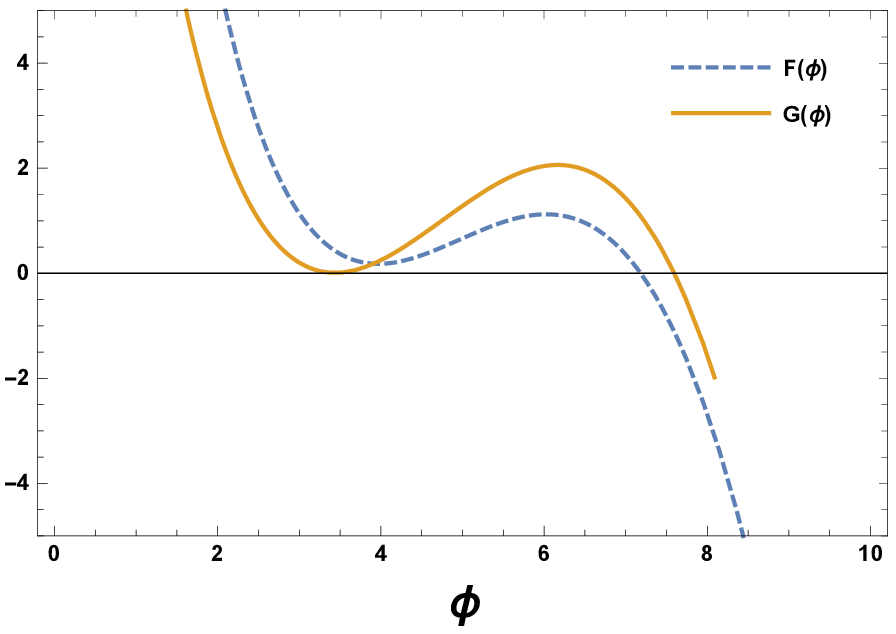}
(b)
\includegraphics[width=.28\textwidth]{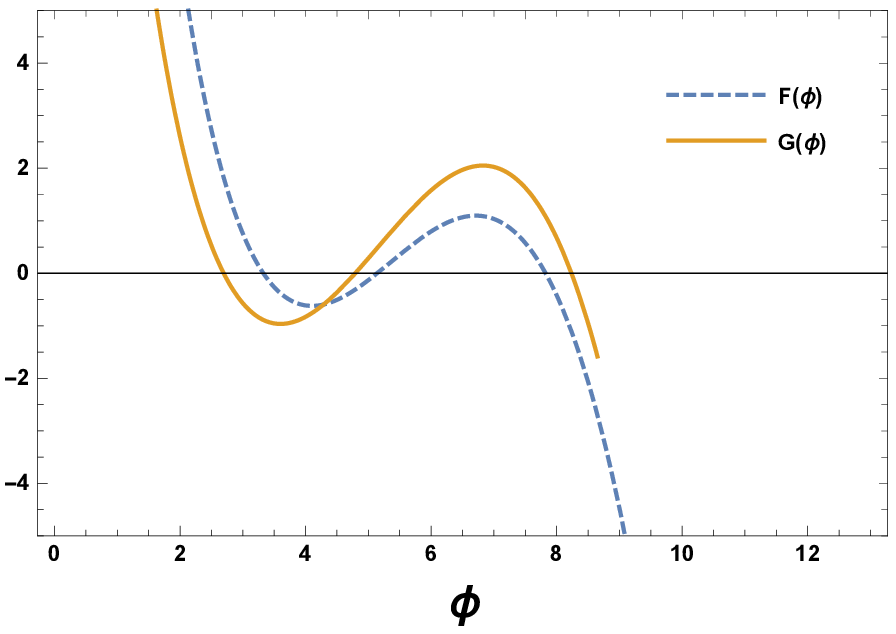}
(c)
\caption{$(a) \phi_0=2 \qquad, \qquad (b) \phi_0=2.37 \qquad and \qquad (c)  \phi_0=2.6$}
\end{figure}
According to the above plots, both functions are positive for different $\phi_0$ in different ranges, and both conjectures are valid in these ranges. As seen in the above figures, for $\phi_0 \leq 2.37$, two curves meet the $\phi$ axis at only one point, and the two conjectures are compatible only in one interval. For example, in figure $2(a)$ in the range of $\phi \leq 6.141$ and for figure $2(b)$ in the range of $\phi \leq 7.184$, the two conjectures are compatible.
If for $\phi_0>2.37$, two curves intersect the $\phi$ axis at three points, in this case, these two conjectures are compatible in the two regions. For example, in figure $2(c)$, the two conjectures are not compatible in the intervals of $\phi \leq 2.701$ and $5.142 \leq \phi \leq 7.822$. We can also find that when $\phi_0$ takes different values, we have various allowed ranges and regions.
\begin{figure}[hbt!]
\includegraphics[width=.4\textwidth]{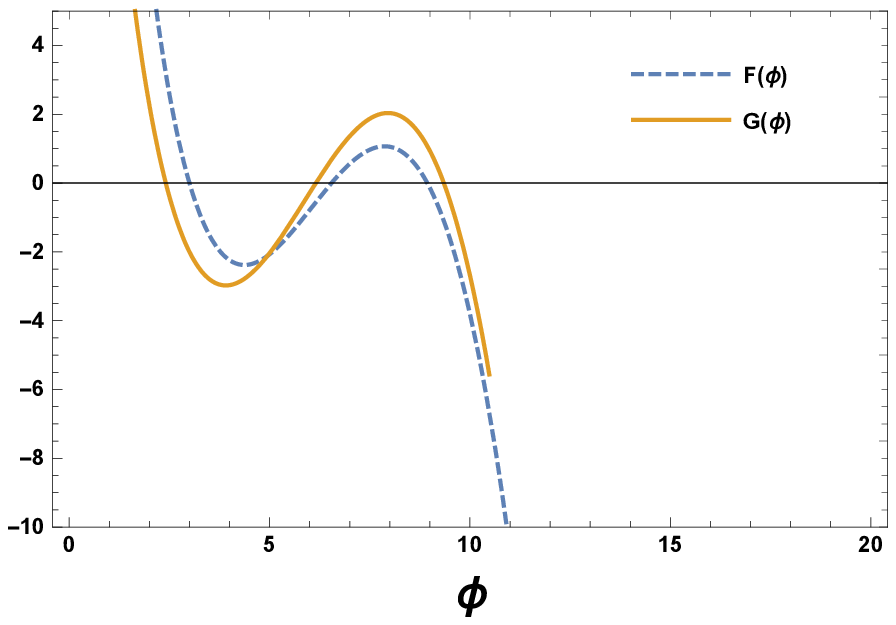}
(a)
\includegraphics[width=.4\textwidth]{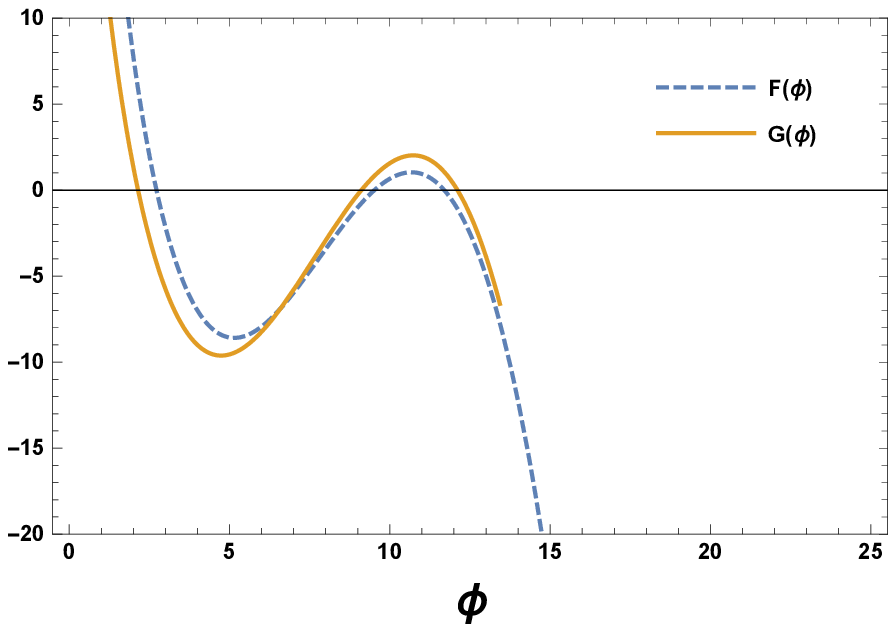}
(b)
\caption{$(a)  \phi_0=3 \qquad, \qquad (b)  \phi_0=4$}
\end{figure}
For $\phi_0=3$, which is shown in figure $3(a)$, two conjectures are satisfied between $\phi \leq 2.394$ and $6.556\leq\phi \leq 8.948$. Also, two conjectures are met for $\phi_0=4$ in the range of $\phi\leq 2.141$ and $9.559\leq \phi \leq 11.675$. The potential structure for the fourth model equation(41) in Einstein's framework is similar to model III in equation (32). Since, the SWGC and SSWGC of the swampland program are proportional to the derivatives of higher orders, as is apparent in the equations (5) and (6).
Therefore, for the fourth model, the same results as model III are obtained, and thus the last two models satisfy the FRDSSC, SWGC, and SSWGC, respectively.
Thus, these two models are more compatible with the swampland program, which is somehow related to quantum gravity. They can be considered desirable models to investigate the universe's evolution.
They are considered favorable inflation models in terms of string theory structure. This way, other cosmological applications of these models can be investigated more profoundly and compared with the latest observable data.

\section{Concluding remarks}
In this article, we want to check four inflation models, such as  composite NJL Inflation(NJLI),   Glueball Inflation(GI),  super Yang-Mills Inflation (SYMI), and Orientifold Inflation (OI), with two conjectures of the swampland program: scalar weak gravity conjecture (SWGC) and strong scalar weak gravity conjecture (SSWGC). Since all these models violate the dS swampland conjecture(DSC) but are compatible with (FRDSSC) through manual adjustment of free parameters of the mentioned conjecture. We studied the simultaneous compatibility of each model with these two new conjectures. Despite being consistent with (FRDSSC), we found that all models are not compatible with the other conjectures of the Swampland program in all regions, and these conjectures are only satisfied in a specific area. Also, Due to the presence of constant parameter $(\phi_{0})$ in the higher orders derivatives, the (SYMI) and (OI) among all the models are more compatible with all conjectures of the swampland program. They can provide a more significant amount of satisfaction with all of them. They can be suitable and accurate inflation models for a more profound examination of universe developments. We determined a particular region for these models are compatible with (FRDSSC), (SWGC), and (SSWGC) simultaneously.\\
We can ask some questions, such as What are the consequences of these conjectures for other models and theories? are there inflationary models that satisfy all conjectures in all regions? Is it possible to extend these conjectures to be consistent with any of the inflationary theories be compatible? We have left the examination of these questions to future work.\\\\
Conflict of Interest\\
The authors declare that they have no known competing financial interests or personal relationships that could have appeared to influence the work reported in this paper.\\
Data Availability Statement\\
Data sharing is not applicable to this article as no datasets were generated or analyzed during
the current study.\\


\begin{thebibliography}{11}
\bibitem{a}
Yuennan, J.; Channuie, Ph. Composite Inflation and further refining dS swampland conjecture. arXiv 2022, arXiv:2208.09842.
\bibitem{b}
Vafa, C. The string landscape and the swampland, arXiv 2005, arXiv:hep-th/0509212.
\bibitem{c}
Kadota, K.; Shin, C. S.; Terada, T.; Tumurtushaa, G. Trans-Planckian censorship and single-field inflaton potential. Journal of Cosmology and Astroparticle Physics 2020, 2020, 008.
\bibitem{d}
Oikonomou, V. K. Rescaled Einstein-Hilbert gravity from f(R) gravity: Inflation, dark energy, and the swampland criteria. Physical Review D 2021, 103, 124028.
\bibitem{e}
Ooguri, H.;  Vafa, C. On the Geometry of the String Landscape and the Swampland. Nuclear physics B 2007, 766, 21.
\bibitem{f}
Trivedi, O.  Rejuvenating the hope of a swampland consistent inflated multiverse with tachyonic inflation in the high-energy RS-II braneworld. Modern Physics Letters A 2022, 37, 2250162.
\bibitem{h}
Das, S. Distance, de Sitter and trans-Planckian censorship conjectures: The status quo of Warm inflation. Physics of the Dark Universe 2020, 27, 100432.
\bibitem{i}
Arkani-Hamed, N.;  Motl, L.;  Nicolis, A.;  Vafa, C. The string landscape, black holes and gravity as the weakest force. Journal of High Energy Physics, 2007, 2007, 060.
\bibitem{j}
Mohammadi, A.; Golanbari, T.; Saaidi, K. Beta-function formalism for k-essence constant-roll inflation. Physics of the Dark Universe 2020, 28, 100505.
\bibitem{k}
Sadeghi, J.; Pourhassan, B.; Gashti, S. N.; Upadhyay, S. Swampland conjecture and inflation model from brane perspective. Physica Scripta 2021, 96, 125317.
\bibitem{l}
Sadeghi, J.;  Pourhassan, B.; Gashti,  S. N.; Upadhyay,  S. Weak gravity conjecture, black branes and violations of universal thermodynamics relation. Annals of Physics 2022, 447, 168168.
\bibitem{m}
Shokri, M.; Sadeghi, J.; Herrera, R.;  Gashti, S. N. Warm inflation with bulk viscous pressure for different solutions of an anisotropic universe. arXiv 2021, arXiv:2112.12309.
\bibitem{n}
Gashti, S. N. Two-field inflationary model and swampland de Sitter conjecture. Journal of Holography Applications in Physics 2022, 2, 13.
\bibitem{o}
Kinney, W. H. Eternal Inflation and the Refined Swampland Conjecture. Physical review letters 2019, 122, 081302.
\bibitem{p}
Yu, T. Y.; Wen, W. Y. Cosmic censorship and Weak Gravity Conjecture in the Einstein–Maxwell-dilaton theory. Physics Letters B 208, 781, 713.
\bibitem{q}
Sadeghi, J.; Mezerji, E. N.;   Gashti, S.N. Study of some cosmological parameters in logarithmic corrected f(R) gravitational model with swampland conjectures. Modern Physics Letters A 2021, 36, 2150027.
\bibitem{r}
Silk, J.; Cassé, M. Swampland Revisited. Foundations of Physics 2022, 52, 86.
\bibitem{s}
 Sadeghi, J.; Gashti, S. N.; Mezerji, E. N. The investigation of universal relation between corrections to entropy and extremality bounds with verification WGC. Physics of the Dark Universe 2021 30, 100626.
\bibitem{t}
Montero, M.; Vafa, C.; Valenzuela, I. The Dark Dimension and the Swampland. arXiv 2022. arXiv:2205.12293.
\bibitem{tt}
Mezerji, E. N.; Sadeghi, J.;  Pourhassan, B. The effect of the WGC condition on the maximal energy extracted from black holes. The European Physical Journal Plus 2022, 137, 1145.
\bibitem{ttt}
Mezerji, E. N.; Sadeghi, J. The correlation of WGC and hydrodynamics bound with $R^4$ correction in the charged $AdS_{d+2}$ black brane. Nuclear Physics B 2022, 981, 115858.
\bibitem{u}
Sadeghi, J.; Gashti, S. N. Anisotropic constant-roll inflation with noncommutative model and swampland conjectures. Eur. Phys. J. C 2021, 81, 301.
\bibitem{v}
Montero, M.; Muñoz, J. B.; Obied, G. Swampland Bounds on Dark Sectors. arXiv 2022. arXiv:2207.09448.
\bibitem{w}
Gashti, S. N.; Sadeghi, J.; Pourhassan,  B. Pleasant behavior of swampland conjectures in the face of specific inflationary models. Astroparticle Physics 2022, 139, 102703.
\bibitem{x}
Anchordoqui, L. ; Antoniadis, I.; Lust, D.  Dark dimension, the swampland, and the dark matter fraction composed of primordial black holes. Phys. Rev. D 2022, 106, 086001.
\bibitem{y}
Sadeghi, J.; Gashti, S. N. Investigating the logarithmic form of f(R) gravity model from brane perspective and swampland criteria. Pramana 2021, 95, 198.
\bibitem{z}
Alvarez-García, R.; Blumenhagen, R.; Kneissl, Ch.;  Makridou, A.; Schlechter, L. Swampland conjectures for an almost topological gravity theory. Phys. Lett. B 2022, 825, 136861.
\bibitem{aa}
Gashti, S.N.; Sadeghi, J. Constraints on cosmological parameters in light of the scalar–tensor theory of gravity and swampland conjectures. Modern Physics Letters A 2022, 37, 2250110.
\bibitem{bb}
Gonzalo, E.; Ibáñez, L. E.; Valenzuela, I. Swampland constraints on neutrino masses. Journal of High Energy Physics 2022, 2022, 88.
\bibitem{cc}
Gashti, S.N.; Sadeghi, J.; Upadhyay, S.; Alipour, M. R. Swampland dS conjecture in mimetic f(R, T) gravity. Communications in Theoretical Physics 2022, 74, 085402.
\bibitem{dd}
Conlon, J. P.; Ning,  S.; Revello, F. Exploring the holographic Swampland. Journal of High Energy Physics 2022, 2022, 117.
\bibitem{ee}
Sadeghi, J.;  Gashti, S.N.; Darabi, F. Swampland conjectures in hybrid metric-Palatini gravity. Physics of the Dark Universe 2022, 37, 101090.
\bibitem{ff}
Hamada Y.; Montero, M.; Vafa, C.; Valenzuela, I. Finiteness and the swampland. J. Phys. A: Math. Theor 2022, 55, 224005.
\bibitem{gg}
Gashti, S.N.; Sadeghi, J. Inflation, swampland and landscape. International Journal of Modern Physics A 2022, 37, 2250006.
\bibitem{hh}
Mishra, R.K. Confinement in de Sitter Space and the Swampland. arXiv 2022. arXiv:2207.12364.
\bibitem{ii}
Gashti, S.N.; Sadeghi, J. Refined swampland conjecture in warm vector hybrid inflationary scenario. The European Physical Journal Plus 2022, 137, 1.
\bibitem{jj}
Cribiori, N.; Dierigl, M.; Gnecchi, A.; Lust, D.; Scalisi, M. Large and Small Non-extremal Black Holes, Thermodynamic Dualities, and the Swampland. arXiv 2022. arXiv:2202.04657.
\bibitem{kk}
Shokri, M.;  Setare, M. R.; Capozziello, S.; Sadeghi, J. Constant-roll f(R) inflation compared with cosmic microwave background anisotropies and swampland criteria. The European Physical Journal Plus 2022, 137, 639.
\bibitem{ll}
Sadeghi, J.; Shokri, M.;  Alipour, M. R.;  Gashti, S.N.  Weak Gravity Conjecture from Conformal Field Theory: A Challenge from Hyperscaling Violating and Kerr-Newman-AdS Black Holes. Chinese Physics C 2022, 47, 015103.
\bibitem{100}
Channuie, P.; Xiong, C. Unified composite scenario for inflation and dark matter in the Nambu–Jona-Lasinio model. Phys. Rev. D 2017, 95, 043521.
\bibitem{101}
Samart, D., Pongkitivanichkul, C.;  Channuie, P. Composite dynamics and cosmology: inflation. Eur. Phys. J. Spec. Top. 2022, 231, 1325.
\bibitem{102}
Channuie, P.; Jorgensen, J. J.;  Sannino, F.  Composite inflation from super Yang-Mills theory, orientifold, and one-flavor QCD. Phys. Rev. D 2012, 86 , 125035.
\bibitem{103}
Bezrukov, F.; Channuie,  P.; Joergensen, J. J.; Sannino, F. Composite inflation setup and glueball inflation. Phys. Rev. D 2012, 86 , 063513.
\bibitem{104}
Feo, A.;  Merlatti, P.; Sannino, F. Information on the Super Yang-Mills spectrum. Phys. Rev. D 2004, 70, 096004.
\bibitem{1}
Obied, G.;  Ooguri, H.; Spodyneiko, L.;  Vafa, C. De Sitter Space and the Swampland. arXiv 2018. arXiv:1806.08362.
\bibitem{2}
Andriot, D.; Roupec,  Ch. Further Refining the de Sitter Swampland Conjecture. Fortschritte der Physik 2019, 67, 1800105.
\bibitem{3}
Liu, Y. Higgs inflation and its extensions and the further refining dS swampland conjecture. Eur. Phys. J. C 2021, 81, 1122.
\bibitem{4}
Garg, S. K.; Krishnan, C.  Bounds on slow roll and the de Sitter Swampland. JHEP 2019, 11, 075.
\bibitem{5}
Ooguri, H.; Palti, E.; Shiu, G.; Vafa, C. Distance and de Sitter conjectures on the Swampland. Phys. Lett. B 2019, 788, 180.
\bibitem{6}
Palti, E. The weak gravity conjecture and scalar fields. JHEP 2017, 08, 034.
\bibitem{7}
Palti, E. The Swampland: Introduction and Review. Fortschritte der Physik 2019, 67, 1900037.
\bibitem{8}
Gonzalo, E.; Ibáñez, L.  A Strong Scalar Weak Gravity Conjecture and some implications. JHEP 2019, 08, 118.
\end{thebibliography}
\end{document}